\documentclass[review, 12pt]{elsarticle}

\usepackage{amsmath}
\usepackage{url}
\usepackage{graphicx}
\usepackage{tabularx}
\usepackage{ragged2e}
\usepackage{rotating}
\usepackage{fullpage}
\usepackage{upgreek}
\usepackage{natbib}

\newcolumntype{Y}{>{\RaggedRight\arraybackslash}X}
\newcommand{\AG}{\textsc{AgesGalore2}}

\begin{document}

\begin{frontmatter}

\title{Single-grain dose-distribution measurements by optically stimulated luminescence using an integrated EMCCD-based system}

\author[label1]{Greilich, S.\corref{cor1}}
\ead{s.greilich@dkfz.de}
\author[label2]{Gribenski, N.}
\author[label3,label4]{Mittelstra{\ss}, D.}
\author[label3]{Dornich, K.}
\author[label5]{Huot, S.}
\author[label2,label6]{Preusser, F.}

\address[label1]{Department of Medical Physics in Radiation Oncology, German Cancer Research Center (DKFZ), Im Neuenheimer Feld 280, D-69120 Heidelberg, Germany}
\address[label2]{Department of Physical Geography and Quaternary Geology, Stockholm University, 10691 Stockholm, Sweden}
\address[label3]{Freiberg Instruments, Delfter Stra{\ss}e 6, 09599 Freiberg, Germany}
\address[label4]{Department of Geography, Justus-Liebig-University, Senckenbergstr. 1, 35390 Gießen, Germany}
\address[label5]{D\'epartement des sciences de la Terre et de l'atmosph\`ere, Universit\'e du Qu\'ebec \`a Montr\'eal, CP 8888 Succ. Centre-Ville, Montr\'eal, Qu\'ebec, Canada H3C 3P8}
\address[label6]{Present address: Institute of Earth and Environmental Sciences - Geology, University of Freiburg, Albert Stra{\ss}e
23-B, 79104 Freiburg, Germany}

\cortext[cor1]{Corresponding author, Tel: +49-(0)6221-42-2632, Fax: +49-(0)6221-42-2665}

\begin{abstract}
We report on the feasibility of assessing single-grain dose-distributions by using an EMCCD-based imaging system with complementary analysis software. Automated image-processing was successfully applied to compensate sample motion and for automated grain identification. Following a dose recovery test, 74\,\% of the grains were recognized successfully, and 44\,\% exhibited a suitable OSL dose response behavior to interpolate an equivalent dose value and a central dose recovery ratio of 1.038 was obtained.
\end{abstract}


\end{frontmatter}

\section{Introduction}
Luminescence dating is based on the measurement of a signal emitted by minerals such as quartz and feldspar during light stimulation. Traditionally, this signal is detected by a photomultiplier tube (PMT), collecting without distinction photons expelled by several hundreds to thousands grains at the same time. Individual grains of a sample can, however, differ in terms of their initial and overall luminescence emission output and behavior. This can be due to variation in radiation sensitivity and saturation characteristics but also partial bleaching, multiple events or post-deposition disturbance recorded in the sample. When analyzing a signal integrated from several grains , this essential information is lost and wrong age estimates can result \cite{Arnold2012}. The determination of equivalent dose distribution on a single grain level can, in contrast, reveal the presence of partial bleaching or mixed grain populations \cite{Olley1999}, allowing to extract sub-population of interest by using statistical tools such as the Minimum Age Model \cite{Galbraith1999} or the Finite Mixture Model \cite{Galbraith1990}. 
The first attempts of single grain equivalent dose measurements using OSL were carried out by Lamothe et al. \cite{Lamothe1994} and Murray and Roberts \cite{Murray1997}, who hand-picked each grains on individual aliquots and measured  them one by one; a laborious and time consuming procedure. The approaches that have been pursued subsequently can be grouped into two principal strategies: Either the entire sample is stimulated simultaneously and a detector records the spatial luminescence intensity distribution (“imaging”) or confined parts of the sample are stimulated subsequently and any suitable photon counting device can be used for OSL detection (“scanning”). Both methods have to provide the accurate assignment of OSL intensity to a specific grain, which uncertainties in the definition of the area of signal integration, signal cross talk between adjacent grains, and variability in mechanical position of the sample during repeated irradiation and readout militate against.

The most widespread single grain system, an automated \textsc{Ris{\o}} reader with additional laser-scanning head \cite{Duller1999}, uses sample holders with 100 holes to host sunk-in single grains and a system of landmarks. This allows for assignment of the OSL intensity to individual grains. However, laser position, power fluctuation and non-uniform irradiation can yield variation in recorded luminescence which is not due to the respective grain properties \cite{Ballarini2006}. Also, grains have to be removed from the original context, which for example deteriorates potential information on small-scale dose-rate variation \cite{Rufer2009}.

More recently and with the advent of affordable, highly sensitive charged coupled devices (CCDs) and electron multiplier CCD (EMCCDs) it was shown that luminescence dating of stone surfaces \cite{Greilich2002, Greilich2006b} and sediment samples \cite{Pfeifer2009, Clark2012} based on the imaging approach is within reach. This would allow the simultaneous stimulation of all grains in a sample, avoiding problems of laser scanning, and thus a wide choice of light sources and potential samples (grain size, sample preserved in their original context etc.).

To constitute a viable alternative in luminescence dating with a sufficiently large number of grains, a process for automated identification (“segmentation”) of random grains and compensation of misalignment (“registration”) has to be accessible which is not yet the case. In this article, we therefore report on the feasibility of routine single grain OSL dating by using a novel, EMCCD-based, integrated system with appropriate image-processing. We believe that three major issues have to be solved to implement the method: sample motion, automated grain identification and signal cross-talk, where we investigate the first two while the latter will be subject of a separate study.

\section{Equipment}

\subsection{Luminescence Reader}
\begin{figure}[!htbp]%
\includegraphics[width=\columnwidth]{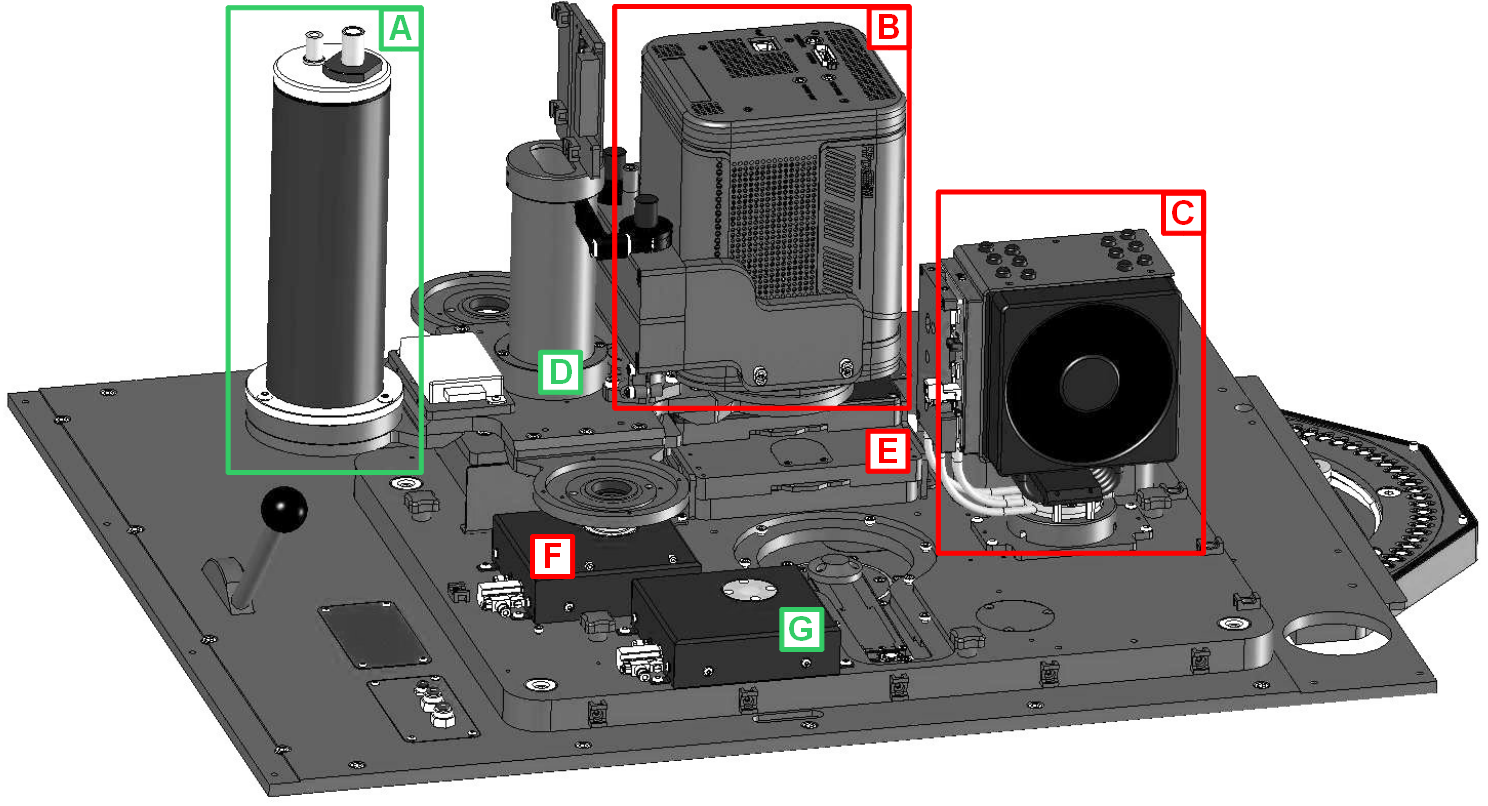}%
\caption{Schematic build-up of the \textsc{Lexsyg re007 "Delorean"} detection and stimulation unit. Green framed units were not used in the experiments: A: ET Enterprises 9235Q photo multiplier. B: Princeton Instruments ProEM512B eXcelon camera. C: solar simulator. D: detector changer. E: multi-wavelength OSL/IRSL-stimulation unit with two 6x filter wheels attached. F: 1.95 GBq high homogeneous ring $\upbeta$-source. G: 1.85 GBq high dose rate overhead $\upbeta$-source. }%
\label{fig:Fig1}%
\end{figure}

All measurements were performed with a \textsc{Freiberg Instruments Lexsyg} research luminescence reader (Fig. \ref{fig:Fig1}), built in 2012 and nicknamed \textsc{Delorean}. In addition to a UV-Vis- and a Vis-NIR-photomultiplier tube mounted besides the used EMCCD camera on a detector changer, the device is equipped with a ‘solar simulator’ consisting of a multi-wavelength power-LED-array \cite{Richter2013} used for bleaching the sample before experiments. For dose regeneration with $\upbeta$-radiation a $^{90}$Sr/$^{90}$Y ring source was used. It delivers a dose rate of about 0.0511\,Gy$\cdot$s$^{-1}$, with a variation of less than 5\,\% over the aliquot area \cite{Richter2012}.

\subsection{Detector}

To perform spatially resolved measurements, \textsc{Delorean} is equipped with a \textsc{Princeton Instruments} ProEM512B eXcelon EMCCD camera (ex works, \cite{Richter2013}). The camera features a back–illuminated, UV-fluorescence-coated e2v CCD97 sensor (512$\times$512 pixels, 16x16 $\upmu$m$^2$) which generates images of 16-bit signal depth. The sensor arrays quantum efficiency is approximately 45\,\% (380\,nm UV quartz emission), 70\,\% (410\,nm feldspars emission), or 85-95\,\% (visible range), respectively \cite{excelon12}. The device was operated in Electron Multiplier (EM) mode. In this mode, the photoelectrons induced by incoming photons are multiplicated before electron-to-count conversion by a high voltage gain register. This increases the signal level in proportion to the read out noise and therefore enhance the signal-to-noise ration in low light applications \cite{TechNote14}.

\subsection{Optics}

The optical setup comprises a lens for light collection built into the OSL stimulation unit and independent of the detector chosen. Two additional lenses attached directly to the camera, focus the image onto the CCD. All lenses are spherical, made of fused silica and have UV/Vis antireflection coating. The setup was calculated to have a numerical aperture (NA) of 0.526, a transmission of $\approx$97\,\% (at 365$\pm$25\,nm), and a magnification of 0.89 (mapping value 18\,$\upmu$m per pixel). The high NA spherical lenses cause considerable spherical aberration. The experimental obtained Michelson fringe visibility (image contrast) of an interference pattern (2 line-pairs / mm, corresponding to a 250\,$\upmu$m structure) is about 30\,\% in the inner 4\,mm image area and decreases further towards the field-of-view edges due to the secondary effect of astigmatism. Practically, this leads to blurry images and halos around a luminescent grain (Fig. \ref{fig:Fig5}). On the other hand the setup has the advantage of increase depth of field, minimizing the effects of chromatic aberrations or sample height variation and thus dispensing the need for refocusing with respect to the detection wavelength or sample.

\subsection{Sample discs and reflective images}

Preliminary tests showed the necessity of obtaining reflective light images (photos) after each record. On the one hand, they are used to correct the positioning uncertainty of the sample arm (see \ref{sec:img_align}). Also they serve as reference pictures for the grain detection algorithm (see \ref{sec:ROI}). Therefore, special stainless steel sample discs of 10\,mm diameter with mirroring surface were developed. They had been polished to remove potential stray light by reflecting the angular irradiating stimulation light to the opposite direction and not towards the camera optics.

\subsection{Stimulation light and detection filters}

For the experiments, blue stimulation light with an intensity of 60\,mW/cm$^2$ at the sample was used, achieved by a ring five power LEDs narrowed to 458$\pm$5\,nm by filters \cite{Richter2013}. For OSL, the detection window was set from 340\,nm to 390\,nm (2.5\,mm \textsc{Hoya} U340 and a \textsc{Delta} BP 365/50 EX interference filter). For reflected light images, no color filters were employed but overexposure was prevented by a ND40B neutral density filter (\textsc{ThorLabs}, optical density 4.0) and an additional aperture with 8.5 mm inner diameter limited spherical aberration and reflection caused by stimulation light flares.

\subsection{Measurement software}

The sequences were built using \textsc{LexStudio2} version 1.0.6c (January 2014) with the new developed "photo" sequence command to take reflective images after each OSL measurement. The blue LED stimulation ($458\pm 5$\,nm) was used as flash light. The image acquisition was performed with \textsc{Princeton Instruments WinView} version 2.6.18 (September 2013). Image exposure was triggered automatically by \textsc{LexStudio2} using the cameras external synchronization port. For data evaluation, \textsc{AgesGalore2GUI} version 0.1.0.3 (March 2014) was applied, using \textsc{ImageJ} version 1.48r (developers build, February 2014) and \textsc{AgesGalore2} build 746 (February 2014) as function libraries.

\section{Image and data processing}

\begin{figure}[!htbp]%
\includegraphics[width=0.7 \columnwidth]{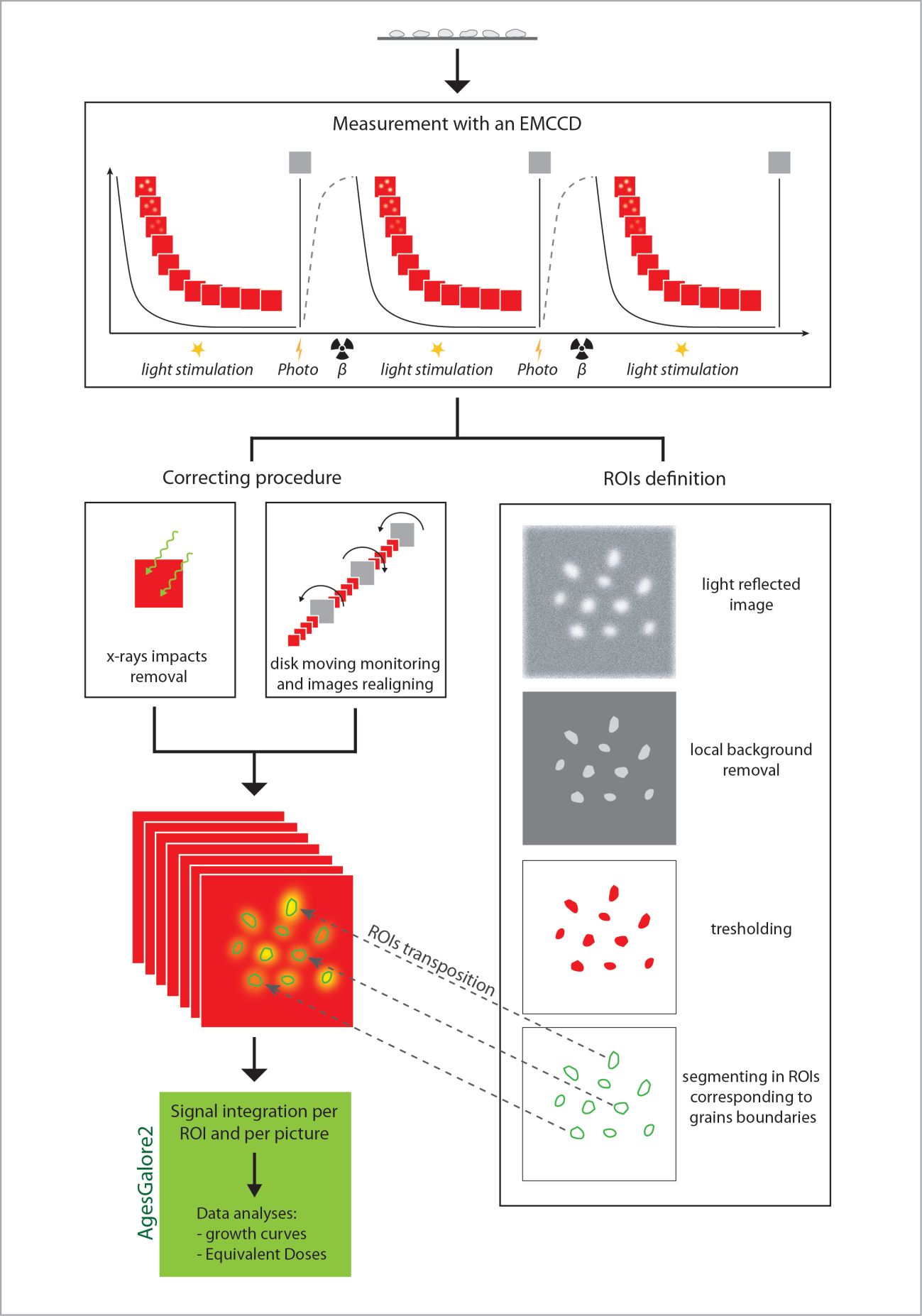}%
\caption{Workflow of single grain OSL dose recovery tests.}%
\label{fig:Fig2}%
\end{figure}

The methodical work flow of single grain  dose evaluation using an EMCCD detector can be divided into three main steps: (i) image stack correction, (ii) region-of-interest definition per grain, (iii) single grain equivalent dose calculation (Fig. \ref{fig:Fig2}).

To achieve the first and the second step, image processing methods were evaluated. The Public Domain biomedical image processing software ImageJ \cite{Schneider2012} was used as function library and plug-in platform.
The third step, the equivalent dose evaluation, is performed by the free  commando line based luminescence dose calculation software \textsc{AgesGalore2}, a lean redefined successor of \textsc{AgesGalore} \cite{Greilich2006b}.
Both software utilities are embedded in the free graphical interface software \textsc{AgesGalore2GUI}\cite{AG2GUI14}\footnote{Both distributed freely under GNU General Public License v3. They can be used, redistributed and modified freely. The copyrights lay by Steffen Greilich (AgesGalore2) respectively Freiberg Instruments GmbH (AgesGalore2GUI). The copyrights and terms of use of ImageJ and his Plug-ins remain by their owners.}.\\

\textsc{AgesGalore2GUI} serves as input mask for all major algorithm parameters, displays the analysis results and performs the data file and sub-program handling\footnote{The GUI is programmed in C\# using the .NET Framework 4.5 Process Class functionality to control a Java Virtual Machine where \textsc{ImageJ} and \textsc{AgesGalore2} are executed by commando line scripting.}. It reads the \textsc{Freiberg Instruments LexStudio2} sequence record (XSYG file) to register the images. Although it is optimized for \textsc{Freiberg Instruments lexsyg} data, when processing the image registration manually and eventually pre-formatting the image stack file, \textsc{AgesGalore2GUI} can be used to evaluate none-lexsyg CCD luminescence measurements too. For this study the AgesGalore2GUI alpha version 0.1.0.3 (March 2014) was used.

\subsection{Image manipulation}

Preliminary tests have shown, that raw image data are corrupted by the following effects:
\begin{itemize}
	\item Randomly falsified CCD areas by the impact of high energy photons
	\item Translated grain locations by not reproducible aliquot positioning deviation
	\item Signal cross talk of nearby grains by optical aberrations (blurring)
\end{itemize}

The first two issues were solved by applying image processing methods, realized in an \textsc{AgesGalore2GUI} controlled \textsc{ImageJ} macro. The last issue will be investigated in a separate study.

\subsubsection{Cosmic ray and secondary X-ray impact removal}	
The ImageJ provided Outlier removal algorithm can decrease the influence of impacts of secondary x-rays (\textit{bremsstrahlung}) due to the nearby beta source and cosmic rays on the CCD \cite{Ferreira2012}. Those rays cause small areas of a few pixels to have values up three to four orders of magnitude above the OSL signal. If an impact occurs inside a region-of-interest, the related $L_x/T_x$ value would be falsified, distorting the dose response curve of the grain. The algorithm identifies the corrupted pixels by comparing them to the median of the surrounding pixels and replaces them with this median value if they are above a user-set threshold. 

\subsubsection{Image alignment}
\label{sec:img_align}
Over repeated measurements, the positioning uncertainty of the sample arm, though less than 50\,$\upmu$m, combined with sliding and rotating due to the arm velocity, can lead to a displacement of the sample carrier relatively to the detector of up to around 180\,$\upmu$m. Therefore, if the the grains location on the CCD picture moves, grain definition and dose response curve calculation would be corrupted. The TurboReg registration plug-in \cite{Thevenaz1998} identifies both translational and rotational sample displacements between two images. In AG2-GUI, this plugin uses the light reflected pictures, which have been taken after each measurement steps. An ImageJ macro sets the first reflective image of the observed image stack as the reference and sets later reflective images as targets. The TurboReg plug-in minimizes the mean-square difference between reference and target and calculate the rotation and translation parameters from some automatically set landmarks. The macro then applies these parameters not just at the target but also on luminescence images progressed before the targeted reflective images was taken, assuming that no sample arm movement happened between them. In this way, the whole image stack is aligned. To prevent edge effects corrupting the grain definition algorithms, a circular mask is drawn after applying the plug-in\footnote{The size of this mask is calculated from the maximum detected aliquot movement, and pixels of the outer region of the mask are set to the mean pixel value of the inner region, overwriting all invalid image regions.}.

\subsection{Region of Interest definition per grain}
\label{sec:ROI}

\subsubsection{Image segmentation strategy: Grain boundaries as ROI}

\begin{figure}[!htbp]%
\includegraphics[width=0.7 \columnwidth]{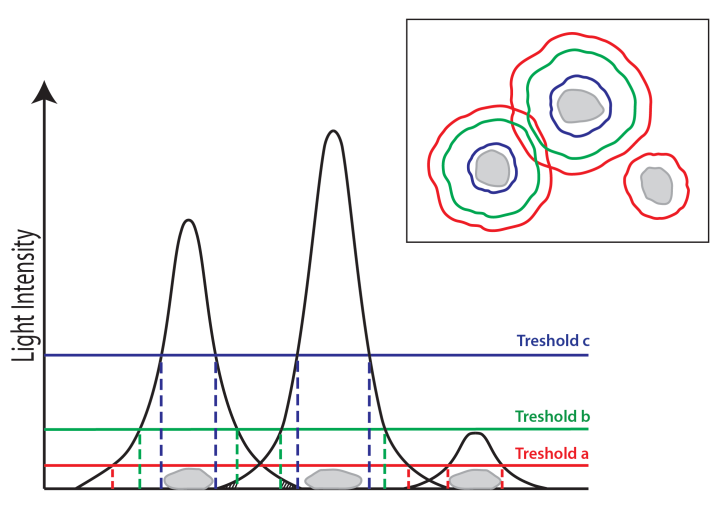}%
\caption{Illustration of thresholding luminescent images. In case, the threshold value is too low (a) the algorithm will not be able to separate the influence sphere per grain in case of two bright grains next close by.  In such a situation, it is necessary to increase the value (b, c), but this will lead to disregard dimer grains.}%
\label{fig:Fig4}%
\end{figure}

\begin{figure}[!htbp]%
\includegraphics[width=0.7 \columnwidth]{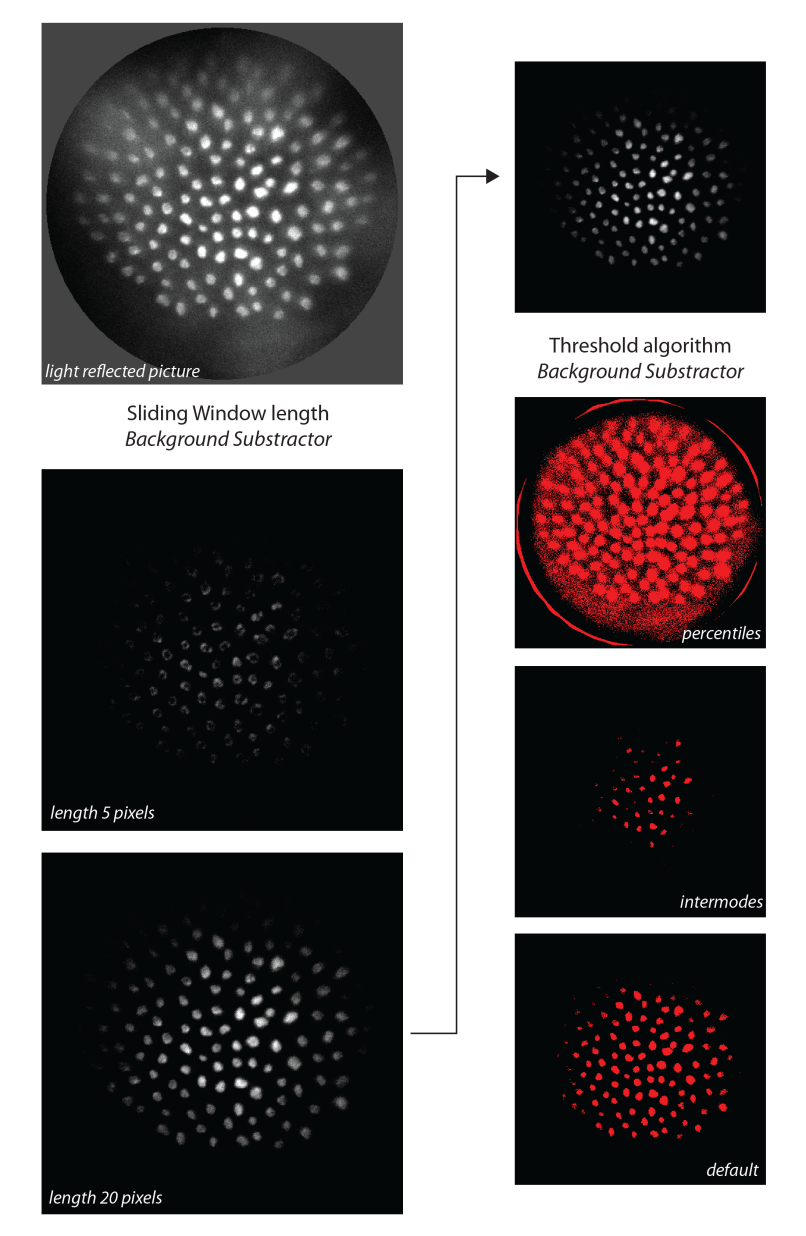}%
\caption{Impact of parameter choice on image quality and segmentation outcome. Left: example of a length too small and an appropriate value for the sliding window side length (Background Substractor). Right: resulting binary image of the “percentiles”, “intermodes” and “default” algorithm for thresholding.}%
\label{fig:Fig5}%
\end{figure}

With the current system, using a Lexsyg reader and an EMCCD camera, digital images records the spatial distribution of luminescence emitted by grains spread over a flat aliquot surface. To attribute a signal per grains, it is therefore necessary to segment the image in regions of interest (ROIs) which will select clusters of pixels per grains.
Two approaches can be applied for segmenting the images. Firstly, focusing on luminescent spot areas and segmenting directly on luminescence images (\textit{a posteriori} segmentation). To isolate areas of bright signal spot emitted by individual grain, a background has to be removed. However, due to the necessity of brightness gradient between the grains, only the brightest grain will be considered while other grains, although bright enough for OSL traditional analysis (SAR, etc.), will be disregarded. In addition, such segmenting approach can yield large ROIs per grain, which are in turn more susceptible to integrate signal emitted by surrounding grains. The second alternative is to define fixed geometrical areas for every grain and to analyze each signal integrated across these areas (\textit{a priori} segmentation). By doing so, one does not aim to include all the pixels exhibiting a luminescence signal, but only those that were found beforehand to belong to a specific grain. Here, we assume that the signal recorded from the area corresponding to the physical grain is not contaminated by signal emitted by surrounding grains (cross talk), as long as a sufficient distance is kept between the grains (Fig. \ref{fig:Fig4}). As this effect is not clearly visible on the luminescent pictures, the identification of the grain boundaries (and whole segmentation procedure) is done using the light reflected pictures of the aliquot.

\subsubsection{ROI segmentation workflow}
Firstly, to automatically identify the grains boundaries, these objects have to be clearly distinguishable from the background, which is inconstant on the light reflected images (reflection light artifacts). The Background Substractor plug-in \cite{Cardinale2010} can mitigate the variable background induced by aberrations and ADC offset. It uses a sliding window histogram approach and takes the most probable value as local background estimate. The side-length of the sliding window has to correspond to (or to be slightly larger as) the diameter of the largest object, i.e. grain 
Then the Threshold plug-in is applied to the pixel values as a first step of grain identification. This approach will transform a grey scale image in a binary image, with one color representing the pixels of interest, and the other the pixels disregarded (background). The plugin provides a choice of numerous algorithms for threshold computing from which will highly depends the binary image resulting (Fig. \ref{fig:Fig5}). Once the image has been correctly segmented, the ImageJ Watershed algorithm \cite{Ferreira2012}recognizes notched areas and dissects them at their narrow point, assuming the areas displayed two overlapping grains. The ImageJ Analyze Particle command converts this into ROIs. It uses geometrical constraints of the objects sought after set by the user, most prominently the circularity and the size ranges.

\subsection{Single-grain equivalent-dose calculation}

\textsc{AgesGalore2} calculates the equivalent dose and the rejection criteria based on an adapted SAR protocol for every pixel and for two-dimensional integration areas (regions-of-interests). The later ability is used to handle every region-of-interests as one quasi-aliquot with one particular dose response curve, evaluated similar to common evaluation methods \cite{Duller2007} including error propagation. \textsc{AgesGalore2} is of similar structure and functionality as its predecessor \textsc{AgesGalore} \cite{Greilich2006a} but had been redefined in a number of aspects:

\begin{itemize}
	\item Based on the Java programming language, it has cross-platform compatibility. Thus, AgesGalore2 could be freed from unnecessary code, e.g. for handling image data, regions of interest, I/O, making the software and its development leaner and faster.
	\item AgesGalore2 was designed as a shell, to be controlled with text commands, scripts or a client/server connection. This allows a tiered structure and to separate graphical user interfaces and data management without intermingling with the functionality.
	\item The software was inherently designed to deal also with coarse grain data while its predecessor was mainly limited to rock slices.
\end{itemize}

A detailed description of the algorithms employed is given in Appendix \ref{sec:detectors}.

\section{Experiments and results}

In order to test our system, and especially the aliquot movement correction, grain boundaries ROI definition and data analysis throughput, we performed dose recovery measurements on quartz samples. Complexities linked to cross-talking are not investigated here and will be the object of a next study, as the sensitivity level evaluation of the system.

\subsection{Samples}
Quartz grains (160-250\,$\upmu$m, standard chemical pre-treatment) were extracted from plunge pool flood sediments from Litchfield National Park, Northern Territories, Australia (May et al., in prep.). According to prior multiple and single grain analyses, these samples exhibit a large number of grains emitting strong and fast component dominated OSL signal.

\subsection{Experimental settings}
Two aliquots with a total of around 250 grains were prepared. To limit signal cross talk, the grains were well separated by checking the disk under a stereo microscope and using a needle. The samples were bleached using the sunlight simulator for 15\,min prior to he actual measurements and bleaching efficiency of the sunlight simulator unit (power of each LED at 50\,\%) was checked by PMT-based OSL readings. A laboratory dose of about 50\,Gy (1000\,s) was chosen for the dose recovery test to prevent any potential difficulties linked to the sensitivity issue (cf protocol tab 1) The Electron Multiplier port of the CCD camera was operated with an avalanche gain of x100, a register readout rate of 5\,MHz, at full-chip resolution and an exposure time of 0.5\,s per image.

\subsection{Rate of success for different image processing parameter settings}
\begin{figure}[!htbp]%
\includegraphics[width=0.4 \columnwidth]{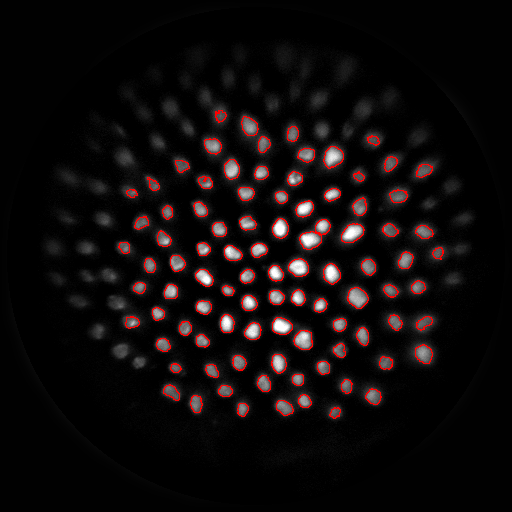}%
\includegraphics[width=0.4 \columnwidth]{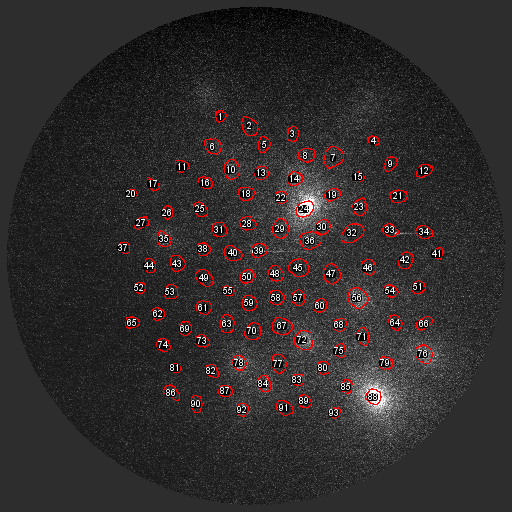}%
\caption{Example of ROI segmentation from the light reflected picture (left) and ROIs transposed over the luminescence picture recording the natural signal (right).}%
\label{fig:Fig6}%
\end{figure}

Four parameters were systematically varied in order to find the most successful combination for the grain segmentation step: the sliding windows side length for the Background Substractor plug-in, the algorithm for the Threshold plug-in, the minimum particle and the minimum circularity for the Analyze Particles tool. 
The best results for the Background Subtractor were achieved using a sliding window length of 20 pixels (equivalent to 360\,$\upmu$m, slightly larger as the average grain diameter). The default thresholding approach based on a variation of the “isodata” algorithm of \cite{Ridler1978} was found to yield best results (Tab. \ref{tab:Tab2}). It calculates the averages of pixels below or equal and above a threshold value test, then computes a new composite average from the two previous mean values, and repeats this process incrementing the threshold value until the latter is above the composite average. The most efficient identification of  grain boundaries has been found using a minimum circularity ($=4\pi\cdot\frac{\text{area}}{\text{perimeter}^2}$) of 0.1 and a minimum size of 50 pixel$^2$ (Fig. \ref{fig:Fig6}).

\subsection{Dose recovery test}
The luminescence intensities were integrated over the 185 ROIs obtained from the best setting above, using the first image (, i.e. the first 0.5\,s) for the initial OSL signal and the last three images (last 1.5\,s) for the background average. The equivalent dose values were calculated based on an exponential curve fitting and a Monte Carlo approach for error calculation. The minimum signal per grain to qualify for equivalent dose evaluation was required to be three standard deviations above background. In total, \textsc{AG2GUI} has been able to calculate an equivalent dose and its associated error for 110 grains (44\,

Without any further exclusion criteria a central dose recover ratio of 1.038$\pm$0.015 and an overdispersion of 13\,\% were found (Fig. 5). The average recuperation ratio of the entire data set was 0.06, with 100\,\% (90\,\%, 50\,\%) of the grains exhibiting a recuperation ratio $<$\,0.15 ($<$\,0.1, $<$\,0.05). Approximately 77\,\% (53\,\%, 30\,\%) of the grains showed a recycling ratio between 0.8 and 1.2 (0.9 and 1.1, 0.95 and 1.05).

Applying a recuperation threshold of 0.05 did not yield significant improvement, both in dose recover ratio (CAM: 1.012$\pm$0.019) and overdispersion (12.1\,\%). In contrast, an additional recycling ratio criterion of 0.1 reduced the overdispersion to 7.6\,\%.

\section{Discussion and conclusion}
The presented approach shows that both the attribution of OSL signal to specific grains and the aliquot movement can be successfully automatized for the needs of single grain dose determination by image processing. Although the success rate of the segmentation procedure is already encouraging, the employment of additional image processing tools (e.g. better thresholding, flat-field correction and deblurring techniques) and/or further development of the optic system should be pursued to allow higher identification rate, even under less ideal (and less artificial) arrangements of grain on the sample disc. 

While the dose recovery test could successfully determine the equivalent dose and a reasonable overdispersion, these experiments do not allow yet assessing the full performance of the system, as this study was performed on well bleached homogeneously irradiated and separated grains, so any potential signal cross talk effects are minimized and not visible in the results. The importance of investigating the impact of this phenomenon in a following study has to be stressed. Even more, as the images are still subject of heavy blurring which most likely enhanced cross talk.

Appropriate signal-to-noise levels seem – at a first glance - to be most challenging when using grains with lower radiation sensitivity and/or lower doses. For imaging devices such as EM-CCDs, the signal per pixel is inevitably lower than that of an integrating device. However, noise cannot only be minimized by longer exposure time, but also lower resolution (e.g. binning per ROI). On the other hand, one is not able to minimize read-out noise beyond a certain limit with these methods. Thus, new expertise in the trade-off of time and spatial resolution has to be sought for this technique. Another solution to increase signal-to-noise ratio would be the adoption of an alternative signal acquisition and progression technique, called high pixel rate photon counting see \cite{Daigle2008, Daigle2012}. While the used camera is technically capable of applying this technique, the practicability remains to be evaluated.

\section*{Author information}
S.G. and N.G. share first authorship.

\section*{Author contribution}
S.G. and N.G. initiated the study and wrote the manuscript, D.M. contributed paragraphs on the lexsyg hardware and AG2GUI. N.G. carried out the measurements. S.G. developed the AG2, D.M. the AG2GUI software and the ImageJ functionality. All authors revised and approved the manuscript.

\section*{Acknowledgements}
We dedicate this article to our late friend and colleague Matthias Krbetschek. His scientific expertise and personal motivation were essential in the development of this method. Special thanks to Steve Grehl and Michael Richter (both Freiberg Instruments) which helped in Programming the AG2GUI. 

\clearpage
\section*{References}
\bibliography{AG2}

\clearpage

\newpage


\newpage

\newpage

\newpage

\newpage
\begin{figure}[!htbp]%
\includegraphics[width=0.9 \columnwidth]{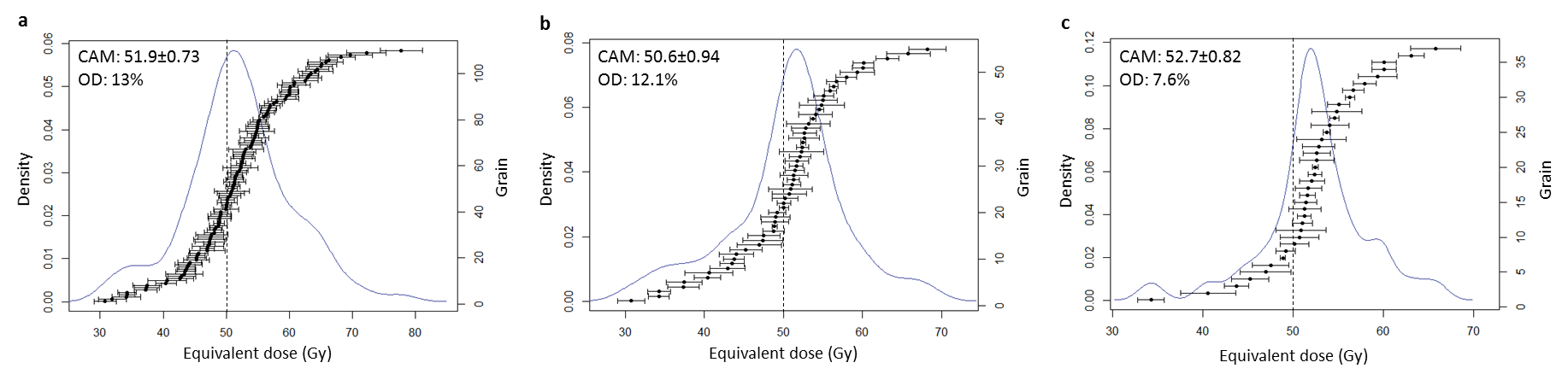}%
\caption{Single-grain equivalent-dose distribution for  a) for the entire data set [UNIT DOSE: GY!!], b) with a recuperation threshold of 0.05 is applied, and c) a recuperation threshold of  0.05 and a recycling criteria between 0.9 and 1.1 applied. }%
\label{fig:Fig7}%
\end{figure}

\clearpage
\begin{table}
		\begin{tabular}{l}
\hline
Dose (artificial nat. 50\,Gy; reg. 25\,Gy, 50\,Gy, 100\,Gy, 0\,Gy, 25\,Gy)\\
\hline
Preheat 230\,°C for 10\,s\\
BlueOSL stimulation 30\,s at 125\,°C\\
Reflected light photo\\
Sunlight simulator bleaching 150\,s\\
\hline
Test Dose (25\,Gy)\\
\hline
Preheat 230\,°C for 10\,s\\
BlueOSL stimulation 30\,s at 125\,°C\\
Reflected light photo\\
Sunlight simulator bleaching 150\,s\\
\hline
\end{tabular}
\caption{DRT protocol \& details}
\label{tab:Tab1}
\end{table}

\clearpage
\begin{table}
\tiny
\begin{tabularx}{\textwidth}{cccccY}

\shortstack{Length window\\BG substractor}& TS algorithm & \shortstack{min size\\particle}&\shortstack{min circularity\\ particle}&\shortstack{ROIs/\\total number of grains}&Matching ROI to grain\\
\hline
5 & Default dark & 50 & 0.1 & 68\,\% &Bad: several ROIs included in one grain area\\
20 & Default dark & 50 & 0.1 & 74\,\% &Good: ROIs defined correspond to grains\\
20 & Intermodes dark & 50 & 0.1 & 115\,\% &Bad: ROIs do not correspond to grains and/or are much larger that grain surfaces\\
20 & Moments dark & 50 & 0.1 & 17\,\% & Good: ROIs defined correspond to grains (but many are ignored)\\
20 & Default dark & 10 & 0.1 & 82\,\% & Mitigate: some ROIs are much smaller than grain surface\\
20 & Default dark & 100 & 0.1 & 62\,\% & Good: ROIs defined correspond to grains (but some ignored as too small)\\
20 & Default dark & 50 & 0.8 & 61\,\% & Good: ROIs defined correspond to grains (but some ignored are not round enough)\\
\end{tabularx}

\caption{Grain identification}
\label{tab:Tab2}
\end{table}

\clearpage
\appendix
\renewcommand*{\thesection}{\Alph{section}}

\section{Data processing in \AG{}}
\label{sec:detectors}

A real-world detector is mimicked in \AG{} as a set of functions $f$ representing its properties to infer the actual number of photons arrivals $\lambda$ at the detector and the corresponding uncertainty\footnote{All uncertainties are reported as \textit{standard uncertainty} ($1\:\sigma$)} $\Delta\lambda$ from the raw digital signal $\tilde{z}$ ("intensity", "counts") that has been recorded. We follow the nomenclature used in \cite{Greilich2006a}. The raw signal ("counts", "intensity") can be described as a series of data points indexed by $n$ taken over several periods of time $m$:

\begin{equation}
\tilde{z}_{n,m}
\end{equation}

In case of PMT data $n=1$ and $m$ describes the "channels" (time intervals). For image-based luminescence data, $n$ runs either over the pixels of an entire image OR the pixels that have been defined as a region of interest (ROI, e.g. a single grain) -- while $m$ runs over the image frames taken subsequently as a series of time intervals, corresponding to the \textit{channels} in the PMT case. The raw signal $\tilde{z}$ will therefore be first summed over a chosen subset of $n$ and $m$, e.g. a ROI and multiple channels:

\begin{equation}
\tilde{z} = \sum_{m=1}^{M}{\sum_{n=1}^{N}{\tilde{z}_{n,m}}}
\label{eq:sumRawSignal}
\end{equation}

The unit of $\tilde{z}$ is assumed to be "counts" (although strickly spoken the quantity is mostly unitless).

\subsection{From raw signal to net signal}
The net signal is derived by (\cite{ProEM2012, TechNote14}):

\begin{equation}
z = \tilde{z} - N\cdot M\cdot (\xi + z_\mathrm{dark} + z_\mathrm{Xray})
\label{eq:signalEMCCD}
\end{equation}

where the notation has been adapted to this document and $\xi$ denotes the bias (background). $z_\mathrm{dark}$ arises due to the thermal creation of additional charge (electrons) and is estimated by
\begin{equation}
z_\mathrm{dark} = n^{e^-}_\mathrm{dark} \cdot \Delta t \cdot G / C
\end{equation}
where $n^{e^-}_\mathrm{dark}$ is the thermal charge generation rate (unit: "number of electrons per second"). For a typical EMCCD camera, $n^{e^-}_\mathrm{dark}$ is approximately $10^{-3}\;e^{-}/s$ at $-70\:C$.\\

$z_\mathrm{Xray}$ is the additional signal due to impact of photons from the build-in radioactive source. Since it was effectively removed by image processing, it was neglected here.\\

\subsection{From net signal to photon arrivals}
The number of photo electrons corresponding to a net signal $z$ can be expressed as

\begin{equation}
n^{e^-} = z \cdot \frac{C}{G}
\end{equation}

and the photon arrivals then be derived from the number of electrons by

\begin{equation}
\lambda = \frac{n^{e^-}}{\eta} = \frac{z}{\eta} \cdot \frac{C}{G}
\end{equation}

where $\eta$ is the quantum efficiency which is here for the sake of simplicity set to unity, i.e.

\begin{equation}
\lambda = z \cdot \frac{C}{G}
\end{equation}

Unphysical, negative photon arrival values can occur at low signals, since $z$ can be negative (see Eq. \ref{eq:signalEMCCD}). Arithmetical operations, e.g. subtraction of two photon arrival values $\lambda_1 - \lambda_2 = \lambda_3$ with $\lambda_3$ being close to zero can strictly spoken be performed in a straight forward manner. While approaches exist to tackle this issue \cite{Greilich2006a}, we neglected this fact in this study for the sake of simplicity and discarded too low signals. 

The uncertainty in photon arrivals is both determined by the shot noise ($\sqrt(\lambda)$) and the electronic read out noise $\sigma_\mathrm{readout}$. $\sigma_\mathrm{readout}$ describes the variation (standard uncertainty) during amplification process and is given in "number of electrons". It is not influenced by $G$. A typical value for an EMCCD system is $2-3\;e^{-}$. 

\begin{equation}
\Delta\lambda = \sqrt{\lambda + N\cdot M\cdot \left(\frac{\sigma_\mathrm{readout}}{\eta}\right)^2} = \sqrt{\lambda + N\cdot M\cdot \sigma_\mathrm{readout}^2}
\end{equation}

To make the photon arrivals comparable if summed over time intervals of different length, the arrival rate $\lambda$ is defined by

\begin{equation}
\Lambda = \frac{\lambda}{M\cdot\Delta t}
\end{equation}

where $\Delta t$ is the length of a time interval (frame, channel) in seconds.\\

\subsection{MASS protocol}
The Multiple Area Single Section (MASS) protocol is an extension of the well-known SAR protocol (\textit{single aliquot regeneration}, \cite{WintleMurray2000}) to multipoint data, e.g. pixels of images, assuming that each data point represents its own independent source of light and dose/age information, respectively. Thus, MASS performs a SAR protocol on each point (yielding e.g. new images of spatial equivalent dose distribution, goodness of fit etc.) \textit{and} on groups of points that have been binned on the level of the raw signal data, i.e. \textit{regions of interest} (ROIs, e.g. for single grains)\footnote{This means especially that -- as described in section \ref{sec:detectors} -- the summed ROI raw signal data first are converted into photon arrivals and then used to establish a growth curve. This is an important difference between \textsc{AgesGalore} and \AG{}. \textsc{AgesGalore} computed equivalent doses for each pixel first and \textit{then} binned/averaged them to a ROI equivalent dose.}. MASS uses the photon arrival rates from multiple ($n>1$) regeneration measurements 

\begin{equation}
\Lambda_\mathrm{reg}(D_i, m_1, m_2)
\end{equation}

where $D_i$ are the regeneration doses (with $i=1...k$, $k>1$ and $D_i \neq D_j$ for at least one combination of $i$ and $j$) and $m_1$, $m_2$ the first and last time interval the raw signal was summed over, respectively, with $m_2\geq m_1$. A "background signal", obtained from later channels / image frames in the shine-down curve is used for correction, so that $L_\mathrm{X}$ is defined as:

\begin{equation}
L_\mathrm{X} = \Lambda^\mathrm{foreground}_\mathrm{reg}(D_i, m_1, m_2) - \Lambda^\mathrm{background}_\mathrm{reg}(D_i, m_3, m_4)
\label{eq:LX2}
\end{equation}

with $m_4\geq m_3 > m_2$. The corresponding test dose data  $\Lambda_\mathrm{reg}^\mathrm{test}(D^\mathrm{test}_i, \tilde{z})$ are used to get $T_\mathrm{X}$:

\begin{equation}
T_\mathrm{X} = \Lambda_\mathrm{reg}^\mathrm{test}(D_i^\mathrm{test}, m_1, m_2) - \Lambda_\mathrm{reg}^\mathrm{test}(D_i^\mathrm{test}, m_3, m_4)
\label{eq:TX}
\end{equation}\\

All $D_i^\mathrm{test}$ have to have the same magnitude. Eventually, $\tilde{L}_\mathrm{X}=L_\mathrm{X}/T_\mathrm{X}$ constitute the points of the dose response (growth curve). 

\subsection{Equivalent dose}
\label{sec:ED}
To compute the equivalent dose $D_\mathrm{eq}$, \AG{} performs curve-fitting to the $\tilde{L}_\mathrm{X}$ values assuming a functional dependency $f_\mathrm{fit}(D, p_i)$ on dose $D$ and a set of $m$ parameters $p_i$. We chose an exponential saturation ($m=3$): 

\begin{equation}
	y = p_1\cdot(1-\mathrm{e}^{-x/p_2}) + p_3
\end{equation}

With the best set of parameters $\hat{p}_i$ found, the natural signal $\tilde{L}_\mathrm{nat}$ is computed in the same way as $\tilde{L}_\mathrm{X}$. $D_\mathrm{eq}$ is  estimated using the model equation based on the reverse of $f_\mathrm{fit}$ and the best set of parameters:

\begin{equation}
D_\mathrm{eq} = f^{-1}_\mathrm{fit}(\tilde{L}_\mathrm{nat}, \hat{p}_i)
\end{equation}

To estimate $\Delta D_\mathrm{eq}$, \AG{} resamples the growth curve elements $\tilde{L}_\mathrm{X}$ and $\tilde{L}_\mathrm{nat}$ according to their corresponding uncertainty and assuming a normal distribution (i.e. parametric bootstrapping). This process is repeated 100 times and the standard deviation of the 100 individual $D_\mathrm{eq}$ results is reported as $\Delta D_\mathrm{eq}$.

\subsection{Test criteria}
\AG computes a number of test criteria partly as given by \cite{WintleMurray2000} and partly specific to the characteristics of pixelized data to check the integrity of the growth curve. This can be done both on a ROI and on the pixel level. The recycling ratio $R$ is computed by dividing $\tilde{L}_\mathrm{1}$ for the lowest (first) regenerated dose $D_1 > 0$) by the recycled signal $\tilde{L}_\mathrm{rec}$

\begin{equation}
R = \frac{\tilde{L}_\mathrm{1}(D_1)}{\tilde{L}_\mathrm{rec}(D_1)}
\end{equation}

The $\tilde{L}_\mathrm{rec}$ is recorded after the regenerated signals and has to have the same dose $D_1$. \AG2 can -- although not intended by \cite{WintleMurray2000} computer multiple recycling ratios if more than one recycled dose exists. The recuperation rate $r$ is defined as the ratio of the last regenerated signal that followed an irradiation with dose 0, $\tilde{L}_\mathrm{X}(0)$, and the natural signal $\tilde{L}_\mathrm{nat}$:

\begin{equation}
r = \frac{\tilde{L}_\mathrm{X}(0)}{\tilde{L}_\mathrm{nat}}
\end{equation}

As a measure of the discrimination between signal and background, \AG{} computes a SNR according to:

\begin{equation}
S = \frac{\tilde{L}_\mathrm{nat}}{\Delta\tilde{L}_\mathrm{nat}}
\end{equation}

\end{document}